\documentstyle[12pt]{article}
\begin{document}
\begin{titlepage}
%{\samepage}
\title{An Analysis of Four-quark Energies\\ in SU(2) Lattice Monte Carlo\\
using the Flux-tube Symmetry}
\author{S. Furui
\thanks{Supported by JSPS and Academy of Finland. e-mail furui@dream.ics.
teikyo-u.ac.jp}\\
School of Science and Engineering,\\
Teikyo University, Utsunomiya 320, Japan
\and A.M.Green \thanks{e-mail green@phcu.helsinki.fi}
and B.Masud\thanks{present address:Department of Physics, Punjab University,
Lahore, Pakistan} \\
Research Institute for Theoretical Physics, \\
P.O.Box 9 SF-00014 University of Helsinki, Finland }

\maketitle
%\date{}

\begin{abstract}
Energies of four-quark systems calculated by the static quenched SU(2)
lattice Monte Carlo method are analyzed in  $2\times 2$ bases for
square, rectangle, tilted rectangle, linear and quadrilateral geometry
configurations and in $3\times 3$ bases for a non-planar geometry
configuration.
For small interquark distances, a lattice effect is taken
into account by considering perimeter dependent terms which are
characterized by the cubic symmetry.
It is then found that a parameter $f$
- that can be identified as a gluon field overlap factor - is rather well
described  by the form $exp(-[b_sE{\cal A}+\sqrt{b_s}F{\cal P}])$, where
${\cal A}$ and ${\cal P}$ are the
area and perimeter mainly defined by the positions of the four quarks, $b_s$
is the string constant in the 2-quark potentials and $E,F$ are constants.
\end{abstract}
\end{titlepage}
\section{Introduction}
QCD is thought to be the basic theory describing the interaction between quarks
and gluons. Unfortunately, until now, because of numerical reasons, this can
only
be checked in systems containing 2 or 3 quarks. Therefore to make the theory
more useful, methods are needed that are much less numerical intensive than
lattice Monte Carlo. These methods should agree with the Monte Carlo
approach for the few quark systems -- the assumption being that their
natural extension to multiquark systems would also give correct results.
Since many of the multiquark systems of interest are derived from hadron-hadron
scattering, in which different quark clusters scatter and rearrange, it is the
$q^2\bar{q}^2$ system that must be first understood.
Therefore, in several attempts to describe meson-meson scatterings from QCD,
the
resonating group method in which a meson is treated as a $q\bar q$ cluster
is a promising approach.  In I=2 $\pi\pi$ and I=1 $K\bar K$
scattering the amplitude is dominated by the direct gluon  and
 quark exchange, and the quark-pair annihilation is expected to play
a minor role. Therefore, as a first step, the detailed analysis of quark
exchange processes from first principles or from  lattice QCD is
desirable.
In a series of papers on the four-quark energies in SU(2) lattice
Monte Carlo\cite{GMP,GMPS}, we considered configurations of two static
$q\bar{q}$ clusters, which we called basis A and the quark interchanged two
$q\bar q$ state basis B. We then analyzed the energy spectra in a
framework similar to the resonating group method using the corresponding
$2\times 2$ bases.  It was observed that the transition potential $v_{AB}$
connecting the bases
needed to be multiplied by a factor $f$ - that can be interpreted as a gluon
field overlap factor - in order that the solution
of the matrix equation in this model  coincided with the two lowest eigen
energies of the Monte Carlo calculation.

Recently the configurations of the four-quark system were extended to
include large square(LS), rectangular(R), tilted rectangle(TR), linear(L),
quadrilateral(Q) and nonplanar(NP) geometries\cite{GMS}.
  An analysis similar to the \cite{GMP,GMPS} showed
that, except for the NP case, the gluon overlap factor $f_0$ derived from the
ground state energy and the $f_1$ derived from the first excited state are
close
to each other, but not identical. In the NP case, the agreement was
worse.

For simplicity of the model, we expect theoretically $f_0=f_1$
and  we tentatively attribute the small difference in $f_0$ and $f_1$
as numerical uncertainties of the Monte Carlo calculation. In the NP case,
however, the analysis in terms of $2\times 2$ bases could, itself,  be an
origin of the difficulty.

In the SU(2) lattice Monte Carlo approach, the transition between different
configurations occurs through multiplications of plaquettes, which
characterize the gluon interaction, with the bases wave function. In the NP
case, some plaquettes are 3-dimensional while in the other five cases
they are purely 2-dimensional. A calculation of the glueball spectrum in
lattice
Monte Carlo shows that the 3-dimensional bent plaquette contains an irreducible
representation of the cubic group which violates the space parity.  In our case
gluon configurations are pinned down by the position of the quarks and
anti-quarks
and the results of the glueball calculation cannot be applied directly, but
evenso we expect that the symmetry of the plaquettes, which makes the
transition from basis A to B could play an important role.

The factor $f$ comes from the flux-tube rearrangement. In the case of TR, Q
and NP, the flux tubes of basis B are not straight lines on the underlying
lattice but they are a superposition of the shortest paths along the links. The
cubic symmetry
of the plaquettes that transform the basis A to the basis B of TR and Q
have different irreducible representations of the cubic group, but in the
NP case, there are two plaquettes that belong to the same irreducible
representation. The excited states of the NP case could be the hybrids i.e.
$q\bar q$ with gluonic excitations.  We, therefore, consider two types of
hybrids $B^+$
and $B^-$ and analyze the NP data in a $3\times 3$ basis using A and $B^\pm$.

In order to extract parameters for use in a resonating group calculation, the
continuum limit, in which the cubic symmetry is replaced by the rotation
symmetry, is to be considered. Thus, parameters which are specific to the cubic
symmetry are not desirable. We, therefore, parametrize the factor
$f$ by a superposition of
an  area dependent term and a perimeter dependent term. When the area is
not a plane surface, the area that we choose is not the sum of the area of the
plaquettes but an approximate minimal surface that is bounded by the links.
The perimeter is kept to be the length of links on the plaquette, since we
want to respect the cubic symmetry which is essential for the parametrization.
We check whether the parameters for different configurations like LS and R
have a common parameter when the two configurations possess the same cubic
symmetry.
The success or failure of this approach is given by the degree to which the
various configurations are described by the same parameters.
Since the gluon overlap factor is not a physical quantity, it could depend
on the lattice size. We expect that the perimeter dependence decreases
as the lattice size becomes larger and in the continuum limit the area
term dominates.

In sect.2 we present the model for fitting the data of the lattice Monte
Carlo, and the numerical results are shown in sect.3. Finally in sect.4
some conclusions are drawn.

%\end{section}
%\newpage
\section{Model}
In fitting the lattice Monte Carlo we could choose three model state
bases\cite{GMP,GMPS}.
\begin{itemize}
\item A: Two links of equal length d that connect $q\bar q$ on 1-3
and on 2-4.
\item B: Two links that connect $q\bar q$ on 1-4 and on 2-3, such that
the number of bending points is a minimum.
\item C: Two links that connect between $q q$ and $\bar q \bar q$ on
1-2 and 3-4, such that the number of bending points is the minimum.
This basis can be regarded as a baryon anti-baryon system in SU(2).
\end{itemize}

As explained in the Appendix of \cite{GMS} the three states are constructed
by contracting the operator
\begin{equation}
T_{ijkl}=P_{ir}(z,x_1)q_r(x_1)P_{js}(z,x_2)q_s(x_2)P_{kt}(z,x_3)q_t(x_3)
P_{lu}(z,x_4)q_u(x_4)\vert 0>
\end{equation}
with tensors
\begin{equation}
A_{ijkl}=\epsilon_{ij}\epsilon_{kl},\quad B_{ijkl}=\epsilon_{il}\epsilon_{kj}
\quad {\rm and} \quad C_{ijkl}=\epsilon_{ik}\epsilon_{jl}.
\end{equation}

Here, in SU(2), the quark field is described by $q_i(x)$ and the antiquark
field by $\overline
{q_i(x)}=-\epsilon_{ij}q_j(x)$ and $P_{ij}(z,x)$ is the colour flux path
from z to x.  Due to the equation $\epsilon_{ik}\epsilon_{jl}=\delta_{ij}
\delta_{kl}-\delta_{il}\delta_{kj}$, the base C is equal to A--B if z can be
chosen to be a common point of the paths of B. This is the case for linear(L),
base $\underline{B_2}$ of quadrilateral(Q) and base $\underline{B_1},
\underline{B_2}$ and
$\underline{B_3}$ of non-planar(NP)(see figs. 4,5 and 6). If there is not a
common
point, $P_{kl}(z,x)$ for B as well as for A should contain a product of links
that starts from z and goes to a quark in B and returns to z, and that starts
from z and goes to a quark in A and returns to z, respectively.
A difference in this expectation value for A and B would make C different
from A--B, but numerical calculation indicates that C is
approximately expressed as a linear combination of A and B, since combinations
A+B, A+C, B+C give very similar energy spectra\cite{GMS}.
In our colour SU(2) static model, we assign two meson states to
A and B, and a baryon anti-baryon state to C.  Therefore, base C is regarded as
redundant and we extract the effective potential in the space A,
in the space B and the transition potential between A and B from the lattice
numerical data.

In the actual calculation the links are fuzzed\cite{PHM}. Here, fuzzing means
replacing a link $U_{ij}$ by a linear combination of $U_{ij}=e^{ig_0A_\mu a}$
and a staple $U_{ik}U_{kl}U_{lj}=e^{ig_0A_\nu a}e^{ig_0A_\mu a}
e^{-ig_0A_\nu a}$, where $\mu\ne \pm \nu$ and $\nu\ne 4$.
It enhances the overlap of the configuration and has the same effect as
tadpole renormalization of the gluon mass.

In the strong coupling approximation of SU(2) gauge theory, the
Wilson matrix for a $q\bar q$ system is
\begin{equation}
<W(L,T,g^2)>\simeq ({1\over 2g^2})^{LT/a^2}=\exp(-b_s LT),
\end{equation}
where L is the length of the link between $q$ and $\bar q$, T is the
Euclidean time, $b_s$ is the string tension and $a$ is the lattice constant.

A generalization to the $q\bar q q\bar q$ system, where the $q$ and
$\bar{q}$ are on the alternate corners of a rectangle of sides
$L_1$ and $L_2$, is\cite{MS,G1}

\[ W=\left[ \begin{array}{cc} e^{-2b_s L_1T} & \epsilon \\
                  \epsilon & e^{-2b_s L_2T} \\ \end{array}\right] \]
where $\epsilon$ is a sum over transitions at time t:
\begin{equation}
\epsilon=-\sum_t e^{-2b_sL_2(T-t)}e^{-b_s L_1L_2}e^{-2b_s L_1t}.
\end{equation}

In Hilbert space the transfer matrix T, which satisfies
\begin{equation}
Z=\int(dU)e^{-S}=TrT^N
\end{equation}
\begin{equation}
S=-{2a\over g_t^2a_0}\sum_{ij({\rm spatial})}ReTr(U_{ij,t+1}^{-1}U_{ij,t})-
{2a_0\over g_s^2a}\sum_{p,t}ReTr(U(p)_t)
\end{equation}
has the matrix element --  see ref.\cite{Cr} page 103 --
\begin{equation}
<U'\vert T\vert U>=exp({2a\over g_t^2a_0}\sum_{ij}ReTr(U_{ij}^{'-1}U_{ij}))
\times exp({2a_0\over g_s^2a}\sum_{p}ReTr(U(p))),
\end{equation}
where $p$ means a plaquette, $a_0$ is the time direction lattice spacing and
$a$ is the space direction lattice spacing. The bare coupling constants
$g_s$ and $g_t$ agree to the lowest order $g_s^2=g_t^2+O(g_t^4)$.

By a group operation g, $U_{ij}$ transforms to $U'_{ij}=gU_{ij}$
and we define unitary operators $R_{ij}(g)\vert U>=\vert U'>$.
In terms of these quantities, T takes the form

\begin{equation}
T=\prod_{ij}(\sum_gR_{ij}(g)exp({2a\over g_t^2a_0}ReTr g))
\times exp({2a_0\over g_s^2a}\sum_{p}ReTrU(p)).
\end{equation}

The area term $\exp(-b_s L_1L_2)$ in eq.(4) originates from the sum of the
spatial plaquettes as they tile the large rectangular
plaquette of  $L_1\times L_2$. Therefore, the dominant part of the transition
potential
between the model space A and the model space B is written in the form
 $v_{AB}exp(-b_sE {\cal A})$  where ${\cal A}$ is the area of
the plaquette that allows transitions between the bases
A and B, and $E$ is to be determined by fitting the Monte Carlo lattice data.

When the plaquette passes through 2 quarks and 2 antiquarks then the degeneracy
of the colour configurations or  the coherence of the products of $U_{ij}$
along the perimeter becomes  important. The gauge transformation $g$ that
connects between the degenerate $Z_2$ bases which are eigen states of
the Cartan subalgebra $t^a\theta^a$, where for SU(2) $t^a=i\sigma^a/2$ and
$t^a\theta^a_\mu$
is proportional to $i\sigma^3 /2\theta$, would be expressed as
a product of the form $exp(t^a\theta^a_\mu x_\mu )$ which is proportional to
the exponential of the length of the perimeter.
 Thus, in addition to the product of the spatial tiles, which can be obtained
by taking $g=1$, we take into account the transition between different $Z_2$
bases which are proportional to the length of the perimeter.

Incorporation of
the perimeter term via renormalization e.g. via the self energy given by the
tadpole diagram is discussed also in the large N expansion\cite{Mak}.
Therefore,
we assume that the potential containing this correction is given in a form
 $v_{AB}exp(-[b_sE {\cal A}+\sqrt{b_s}A_n{\cal P}])$ where ${\cal P}$ is
the length of the perimeter and $\sqrt b_s$ is included  to make $A_n$
dimensionless.  This form has been derived from the case where the four
quarks are at the corners of a rectangle. For this the area ${\cal A}$ and the
perimeter ${\cal P}$ have clear definitions. The further assumption is now made
that,
in general, ${\cal A}$ is the natural \underline{continuum} values --
independent of the underlying lattice structure. This is similar to describing
the two-quark potential on the lattice  as
$V(r)=-e/r+b_sr+v_0$, where $r$ is the \underline{shortest} distance between
the two quarks. It is known that this form is good for $r\ge 2a$. However,
for smaller values of $r$ the lattized form
\begin{equation}
 V(r_{ij})=-(e/r_{ij})_L+ b_sr+v_0
\end{equation}
is neccessary -- see ref. \cite{GMPS}

For the same reasons, those four-quark configurations, where the quarks are
within two lattice spacings, probably contain effects arising from the
underlying lattice. We keep the lattice structure for the perimeter term
${\cal P}$ and take into account this effect partially.

\subsection{Irreducible representations of the cubic group}
We consider now g as an element of the cubic group or the crystallographic
point group. It has the effect of
rotating links on a plaquette and we are interested in group operations which
connect a product of $U_{ij}$ to a product of $U'_{ij}$ which differ
only by a cyclic permutation of the links.
The Wilson loop at a fixed time $\tau$, or the time-local operator, is
characterized by L-tuples ${\cal O}_m(p)=
(\hat f_1,\cdots ,\hat f_L)$ and a class formed by all the L-tuples, which are
identical up to cyclic permutation, is denoted by $[\hat f_1,\cdots ,\hat
f_L]$. The suffix $m$ specifies the orientation of the plaquette and this
numbering follows the definition of \cite{BB}.

Here the $\hat f_j$'s stand for unit vectors in the x-, y- and z- directions:
$e_1, e_2$
and $e_3$\cite{BB}. The charge conjugation transforms
$C[\hat f_1,\cdots ,\hat f_L]=[-\hat f_L,\cdots, -\hat f_1]$.
We define ${\cal O}^r_m(p)=(\hat f_L,\cdots ,\hat f_1)$. In usual SU(N),
C-parity positive and negative bases are given by
$[\hat f_1,\cdots ,\hat f_L]\pm [-\hat f_L,\cdots ,-\hat f_1]$.
In the case of SU(2), there are only C-parity positive states
as the bases.

On the lattice the bases are transformed by the proper rotations
${\cal M}_g$ of the crystallographic point group which are divided into five
classes\cite{Co};
\begin{itemize}
\begin{enumerate}
\item  $E$(identity)
\item  $C_3=\{ C_{3h}, C_{3h}^{-1} \}$ (h=$\alpha,\beta,\gamma,\delta$)
\item  $C_4^2=\{ C_{4h}^2\}$ (h=x,y,z)
\item  $C_4=\{ C_{4h}, C_{4h}^{-1} \}$ (h=x,y,z)
\item  $C_2=\{ C_{2h} \}$ (h=a,b,c,d,e,f)
\end{enumerate}
\end{itemize}

The operations on the bases are given by
${\cal M}_g [\hat f_1,\cdots ,\hat f_L]=[M_g\hat f_1,\cdots ,M_g\hat f_L]$.
The main difference of the classification of the type of the plaquettes
here as compared to other lattice calculations of the glueball spectrum
\cite{BB} is that we fix the
configuration of the links in our model space and so we restrict $M_g$
such that $[M_g\hat f_1, \cdots ,M_g\hat f_L]$ belongs to the class of
$[\hat f_1,\cdots ,\hat f_L]+[-\hat f_L,\cdots ,-\hat f_1]$. Actually, the
link between $q$ and $\bar q$ is fuzzed and we do not know the exact
overlap factor between the states obtained by multiplying a plaquette and
plaquettes  properly rotated which are specified by
$C_k$.  We parametrize the effect of the overlap of the links on the
perimeter of the plaquette by a correction to the above area law dependent
term and search the parameters by fitting the data.

We specify the plaquettes that allow transitions between  the spacial wave
function of A into B in the large square(LS), rectangle(R), tilted
rectangle(TR), linear(L), quadrilateral(Q) and non-planar(NP) geometries as
follows.

\begin{itemize}
\begin{enumerate}
\item LS

The bases and the square plaquette(sp) for the transition, which is
specified as ${\cal O}(sp)_1$, are shown in Fig.1. The suffix 1 corresponds
to the assignment of ref.\cite{BB} for the number of links $l=4$.

The proper rotations that transform ${\cal O}(sp)_1$ into itself
are $C_{2z}, C_{4z}, C_{4z}^{-1}$ and that transform into ${\cal O}(sp)_1^r$
are $C_{2x}, C_{2y}, C_{2a}$ and $C_{2b}$.
\item R

The bases and the rectangular plaquette(rp) for the transition
which is specified as ${\cal O}(rp)_1$ are shown in Fig.2

The proper rotation that transforms ${\cal O}(rp)_1$ into itself
is $C_{2z}$ and that transform into ${\cal O}(rp)^r_1$  are $C_{2x}$ and
$C_{2y}$.

Here it should be emphasized that, in ref.(3), the R-geometry also contains
the four squares with sides $d=1,2,3,4$. These behave as in case LS above.
Since the LS-case has more transformation options than the R-case, it is
expected that squares generate a larger overall interaction -- as is observed
in the lattice data.

\item TR

The bases and the bent plaquettes(bp) for the transition, which is
specified as ${\cal O}(bp)_7$, are shown in Fig.3.  The suffix 7 corresponds to
the assignment of \cite{BB} for the number of links $l=6$. The symmetry
property does not change for larger $l$.

The proper rotation that transforms ${\cal O}(bp)_7$ into itself
is $C_{2c}$ and that transform into ${\cal O}(bp)^r_7$  are $C_{2d}$ and
$C_{2y}$.

\item L

The bases of the linear configuration are shown in Fig.4.
The transition now occurs by the insertion of minimum plaquettes twice at the
same position. We call it the self-energy operation.

\item Q

The bases and the rectangular plaquette(rp) for the transition
which is specified as ${\cal O}(rp)_1$ are shown in Fig.5. The transition to
$B_2$ occurs through the self-energy mechanism as in case L.

The proper rotation that transforms  ${\cal O}(rp)_1$ into itself is
$C_{2z}$ and that transform into ${\cal O}(rp)^r_1$ are $C_{2x}$ and $C_{2y}$.

\item NP

The bases and the plaquettes(p=sp,rp), bent plaquette(bp) and
twisted plaquette(tp) for the transition which are specified as
${\cal O}(p)_1\pm {\cal O}(p)_5$,
${\cal O}(bp)_7\pm {\cal O}(bp)_{12}$ and ${\cal O}(tp)_1$ are
shown in Fig.6.  The suffices 7 and 12 correspond to the assignment of
\cite{BB} for $l=6$. The symmetry property does not change for larger $l$.
The proper rotations among bases ${\cal O}(p)_1$ and ${\cal O}(p)_5$ and
those among bases ${\cal O}(bp)_7$ and ${\cal O}(bp)_{12}$ can be obtained
by the standard method\cite{Co}.

The proper rotations that transform ${\cal O}(tp)_1$ into itself are
$C_{3\beta}, C_{3\beta}^{-1}$ and that transform into ${\cal O}(tp)^r_1$  are
$C_{2a}, C_{2d}$ and $C_{2e}$.

The self-energy mechanism transforms the basis A into the basis $B^+$, which
is created from A by an insertion of ${\cal O}(p)_1+{\cal O}(p)_5$ and
${\cal O}(bp)_7+{\cal O}(bp)_{12}$, but not with $B^-$ which is created from A
 by an insertion of ${\cal O}(p)_1-{\cal O}(p)_5$ and
${\cal O}(bp)_7-{\cal O}(bp)_{12}$.
\end{enumerate}
\end{itemize}
%\end{subsection}
\subsection{The gluon overlap factor}
Among the proper rotations that keep the orientation of the plaquette, the
operation that transforms to ${\cal O}(p)^r$, which is related to the charge
conjugation, would have the main overlap to the basis C.  Typical
examples are obtained by transforming the shortest $\bar q q$ link in the
basis $\underline{B_2}$ of the NP geometry and switching the connection of the
open
circle on the
shortest link to make a $qq$ link and a $\bar q\bar q$ link. Another example
can be produced in the basis $\underline{B_3}$ of NP, but this kind of coupling
of the basis C to the bases A and B occurs only in very restricted
circumstances: a diagram with three paths from a $q$ or a $\bar q$ must
be there and the three paths should be colour neutral.
Therefore we neglect the rotation that transforms ${\cal O}(p)$ to
${\cal O}(p)^r$.

We define the length of the two paths in the state A and the distance
between the two,  in units of the lattice spacing $a$, as $d$ and $r$,
respectively. In the case of L, Q and NP, $r$ is defined as
the distance between quarks 3 and 4.

The strong coupling potential in the direct channel is obtained from
\begin{equation}
<W(L,T,g^2)>=e^{-v(L)T}
\end{equation}
which yields $v(r_{ij})=b_sr_{ij}$.  Including the lattice Coulomb
interaction and the self-energy $v_0$\cite{GMPS}  and defining $v_{ij}$
as (9), we parametrize $v_A=v_{13}+v_{24}, v_B=v_{14}+v_{23}$.

We fit the gluon overlap factor $f$
for LS, R, TR, L and Q by the condition that the solution $\lambda$ of the
equation

\begin{equation}
 det W=det( \left[ \begin{array}{cc} v_A & fv_{AB} \\
                 fv_{AB} & v_B   \\
                 \end{array}\right]
-\lambda\left[ \begin{array}{cc} 1 & f/2 \\
                 f/2 & 1  \\
                 \end{array}\right] ) =0
\end{equation}
agrees with the energy eigenvalues of the lattice calculation\cite{GMS}.
It should be added that the parametrization of the two body potential (9)
is done for each configuration separately and the deviation of the lattice data
from the fitted formula (9) is less than 0.5\%.

In the NP case, we need to analyze the lattice calculation
data using the bases A, $B^+$ and $B^-$. We define the overlap factors
between  A and $B^\pm$ by $f^\pm$.
Since
\begin{eqnarray}
& &<O(p)_i+O(p)_j\vert C_{2k}\vert O(p)_i-O(p)_j>\nonumber\\
&=&-<O(p)_i\vert C_{2k}\vert O(p)_j>+<O(p)_j\vert C_{2k}\vert
O(p)_i>\nonumber\\
&=&-<O(p)_i-O(p)_j\vert C_{2k}\vert O(p)_i+O(p)_j>
\end{eqnarray}
and the matrix W should be Hermitian, we define the
transition matrix between $B^-$ and $B^+$ is proportional to
 the Pauli matrix $-\sigma_y $ multiplied by a factor $t$
and choose $f^\pm$ to be complex.

We fit $f^\pm$ and t from the condition that the solution $\lambda$'s of the
following equation coincide
with lowest three energy eigenvalues of the lattice calculation.

\begin{eqnarray}
 det W=det( \left[ \begin{array}{ccc} v_A & \sqrt 2v_{AB}f^+
           & \sqrt 2v_{AB}f^- \\
            \sqrt 2 v_{AB}f^{+*} & v_B & i t  \\
          \sqrt 2 v_{AB}f^{-*} & -it & v_B \\ \end{array}\right] \\
 -\lambda\left[ \begin{array}{ccc} 1 &\sqrt 2 f^+/2 &\sqrt 2 f^-/2 \\
         \sqrt 2 f^{+*}/2 & 1 & 0  \\
         \sqrt 2 f^{-*}/2 & 0 & 1 \\ \end{array}\right] ) =0
\end{eqnarray}

The matrix elements between A and $B^\pm$ are multiplied by $\sqrt 2$ due to
the normalization of the bases of $B^\pm$.
Algebraic forms of the three $\lambda$'s can be obtained by using Mathematica.
They are functions of $\vert f^+\vert^2+\vert f^-\vert^2$ and
$Im(f^+f^{-*})$. Therefore $f^+$ and $f^-$ are interchangeable.

The explicit forms of $f$ are chosen as follows.
\begin{itemize}
\begin{enumerate}
\item LS

\begin{equation}
f=e^{-[b_sE{\cal A}+\sqrt{b_s}(A_1+A_2){\cal P}]}
\end{equation}
Here $b_s$ is the string tension and ${\cal A}=d^2$  -- the area of the
square.
The parameter $E$ fixes the overlap of the gluon configuration due to the
strong
coupling area dependence. Parameters $A_1$ and $A_2$ fix the
overlap due to the links on the perimeter of the area and ${\cal P}=4d$.
 $A_1$ contains contributions of $C_{2z}$ and $A_2$ contains contributions of
$C_{4z}$ and $C_{4z}^{-1}$.

In principle, one could include an overall normalization factor
$f_c$ in $f$. In fact, if
one omits the data of $(r,d)=(1,1)$ and fits the overlap factor $f$ with
the additional
factor $f_c$, we obtain a much smaller $\chi^2$ with $f_c\simeq 0.87$.
However, in this case $f$ becomes almost
Gaussian and the value of $f$ at $r=0$ and $r=1$ in lattice units becomes very
close to each other.  This means that when we change the lattice spacing
$a$ to a smaller $a'$ the slope of $f$ at  $(r,d)=(1,1)$ and at
$(r,d)=(2,2)$  in the new lattice spacing would be quite different. We compare
the data of $\beta=2.4$ which corresponds to $a=0.12 fm$ and $\beta=2.5$ which
corresponds to $a'=0.082 fm$.  The parameters of the quark-antiquark potential
for the two $\beta$ values are given in Table 1. The overlap factor $f$ of
$(r,d)=(1,1)$ for $\beta=2.5$ turns out to be 0.94.
Thus, we fit $f$ as an exponential and not as Gaussian, and fix the
normalization to be 1. This means that we incorporate the lattice effect
or correction to the area term explicitly into the fitting. We then need to
extrapolate the fitted data to the continuum limit, where the perimeter
term -- except for the self energy term -- would become unimportant.

\item R

\begin{equation}
f=e^{-[b_sE{\cal A}+\sqrt{b_s}A_1{\cal P}]}
\end{equation}
Here ${\cal A}=rd$ and ${\cal P}=2(r+d)$. Since R does not have the symmetry
corresonding to $C_{4z}$ and $C_{4z}^{-1}$, the $A_2$  must be omitted here.
Since the area term dominates over the perimeter term as r and d becomes large,
the
difference in the overlap factor in LS and R becomes small for large
plaquettes.

\item TR

\begin{equation}
f=e^{-[b_sE{\cal A}'+\sqrt{b_s}A_3{\cal P}]}
\end{equation}

We define the projection of the paths 1-4 and 2-3  onto the x-y plane and onto
the  y-z plane by $(x,z)$ -- in  lattice units $a$ . Therefore the length
of the paths in B is $\sqrt{x^2+z^2}=r'$ and
${\cal A}'=dr'$ is the area of the surface that is bounded by the
four links 1-3, 3-2, 2-4 and 4-1. We do not take the sum of the area of the
two plaquettes of Fig.3 since for large $d$ and $r$ it would not
correspond to the continuum limit of interest.
The spectrum for $(d,x,z)=(5,3,4)=(5,4,3)$ becomes very similar to that of
LS with $r=d=5$, in this model.

Although a parametrization specific to the cubic symmetry is not
desirable,  the length of the perimeter is tightly related to the cubic
symmetry of the links: i.e. $A_3$ contains the contribution of $C_{2c}$.
Thus, we take ${\cal P}=2(x+z+d)$ which is the length of the bent plaquette.

\item L

\begin{equation}
f=e^{-\sqrt{b_s}A_0 2r}
\end{equation}

In order to simulate the self energy mechanism we take the perimeter
of the minimum area along the overlapping links of the basis B.
Normalizing $f=1$ for $r=0$, we obtain the above formula.

\item Q

\begin{equation}
f=(e^{-\sqrt{b_s}A_0 2(r-d)}
+e^{-[b_sE{\cal A}'+\sqrt{b_s}A_1{\cal P}]})/2
\end{equation}
where the first term comes from the same mechanism as the linear (L) and the
latter is the rectangular plaquette contribution. The area
${\cal A}'=rd/2$ is the area of the triangle 1-3-4, that is half of the
plaquette area. ${\cal P}=2(r+d)$ is the length of the perimeter of the
rectangular plaquette which transforms under $C_{2z}$.

\item NP

In the nonplanar case, combinations of  bent plaquettes ${\cal O}(bp)_i$
$(i=1,\cdots, 12)$ lead to different parity states\cite{BB}. In our
restricted plaquette bases the symmetric and antisymmetric combinations
of the plaquettes  ${\cal O}(p)_1\pm {\cal O}(p)_5$ and
${\cal O}(bp)_7\pm {\cal O}(bp)_{12}$ lead to  bases $B^+$ and $B^-$
respectively.

 We parametrize $f^\pm=(f_{re}\pm i f_{im})$, where

\begin{eqnarray}
f_{re}=(e^{-\sqrt b_sA_0 2r} +
 2e^{-[b_sE{\cal A}_1+\sqrt{b_s}B_1{\cal P}_1]}
+2e^{-[b_sE{\cal A}_2+\sqrt{b_s}B_2{\cal P}_2]}\nonumber\\
+e^{-[b_sE{\cal A}''+\sqrt{b_s}A_4{\cal P}_3]})/6
\end{eqnarray}
and
\begin{equation}
f_{im}=( e^{-[b_sE{\cal A}_1+\sqrt{b_s}B_1{\cal P}_1]}
+e^{-[b_sE{\cal A}_2+\sqrt{b_s}B_2{\cal P}_2]})/2.
\end{equation}

Here $B_1$ in $f_{re}$ contains the contribution of  $C_{2z}\pm C_{2y}$,
while that in $f_{im}$  the contribution of $C_{4x},C_{4x}^{-1}$
and $C_{2e}$. ${\cal P}_1=2(r+d)$ is
 the perimeter of the plaquette ${\cal O}(p)_1$ and ${\cal O}(p)_5$, and
${\cal A}_1=rd$ is the corresponding area. $B_2$ in $f_{re}$ contains
contributions of $C_{2c}\pm C_{2b}$, while that in $f_{im}$ the
contribution of $C_{3\alpha}, C_{3\alpha}^{-1}$, and $C_{2e}$. ${\cal
P}_2=2(r+2d)$ is the
perimeter of the plaquette ${\cal O}(bp)_7$ and ${\cal O}(p)_{12}$, and
${\cal A}_2=d(r+d)$ is the corresponding area.
$A_4$ contains the contribution of $C_{2c}\pm C_{2b}$ and ${\cal P}_3=2(r+2d)$
 is the perimeter of the plaquette ${\cal O}(tp)_1$. For the corresponding area
${\cal A}''$ we adopted the area of a curved surface
\begin{eqnarray}
{\cal A}''=\int_0^1du\int_0^1dv\vert (u{\bf r}_{13}+(1-u){\bf r}_{42})
\times (v{\bf r}_{23}+(1-v){\bf r}_{41})\vert\nonumber \\
=\int_0^1du\int_0^1dv d\sqrt{r^2(1-2u+2u^2)+d^2v^2}
\end{eqnarray}
which can be calculated analytically.

The ansatz (22) corresponds to an approximation by the regular surface
or an approximation of a surface by a sum of straight lines that connect
between two flux lines in a 3-d space e.g. 23 and 14 in fig.6.
If the four links that surrounds the area are not on a plane, the regular
surface is not necessarily the minimal surface. In fact except for the
plane surface it is only the helicoidal surface which is minimal and
regular\cite{DoC}. A numerical calculation, using the method of \cite{Bus}
shows, however, that the ansatz (22)
is a reasonable approximation especially when $r>>d$. Since the evaluation of
(22) is simple, we adopt this formula for the area of the NP geometry.

The appearance of $t$ in the potential means that the configuration of quarks
and anti-quarks are not enough to classify the energy spectrum and we need
to specify the gluonic degrees of freedom. It is natural in the flux-tube
picture since the links in B are bent and the excited state can be mixed
in. The transition potential connects two different symmetry states which
we specified by $B^+$ and $B^-$.

The appearance of perimeter dependent factors and the splitting
of B into $B^+$ and $B^-$ come from gluonic degrees of freedom, since
the lattice data are not expressed in the bases which contain only the quark
degrees of freedom.
\end{enumerate}
\end{itemize}
%\end{subsection}
%\end{section}

\section{Numerical results}

In order to fit the  parameters $E, A_i$ of LS, R, TR, L, and Q, we tried two
options;
\begin{itemize}
\begin{enumerate}
\item Fit eigen-energies of all the $(r,d)$ sets directly.
\item Calculate the overlap factor $f_0$ from the ground state eigen
energy and $f_1$ from the first excited state eigen-energy by solving
a quadratic equation for a fixed $(r,d)$ set\cite{GMP} and fit the averages of
$f_0$ and $f_1$ of all the $(r,d)$ sets.
\end{enumerate}
\end{itemize}
The two options give slightly different results. The error-bars of the
eigen-energies at small $(r,d)$ are crucial for the relative magnitudes
of the area dependent part and the perimeter dependent part. The error-bar
for small $(r,d)$ is smaller than that for large $(r,d)$, but for $(r,d)=
(1,1)$, there are problems in fitting the two-body potential $v_{ij}$ as
a sum of the three terms given (9)\cite{GMP}. However,  we assume that the
lattice effect is well simulated by the perimeter term, and use
the small error-bars given by the lattice simulation. The results of
the first option i.e. direct eigen-energy fit give following parameters.
\begin{itemize}
\begin{enumerate}
\item LS

We obtain $E=0.296(11)$ and $A_1+A_2=0.080(2)$.

\item R

We fix $E$=0.296 and obtain $A_1=0.057(1)$ which means $A_2=0.023(3)$.

\item TR

We again fix $E$=0.296 and obtain $A_3=-0.091(6)$. In order that the two mesons
does not interact for large $r$ it is necessary that $ lim_{r\to\infty}f=0$.
However, the negative value of $A_3$ does not
cause trouble since the area term dominates in physically reasonable
configurations and $f$ tends to 0 for large $r$.

\item L

We obtain $A_0=0.197(3)$.

\item Q

We fix E=0.296 and $A_0=0.197$ then we find $A_1'=0.051(2)$ which is close to
the value $A_1=0.057(1)$ obtained in R. This is a good indication that our
parametrization is meaningful. When we use the perimeter ${\cal P}=d+r+
\sqrt{r^2+d^2}$ which corresponds to the continuum limit, we obtain for $d=
2,3$ and 4, $A_1'=0.0595$ i.e. closer to $A_1$. We insist, however, on
using the ${\cal P}$ that respects the cubic symmetry.

\item NP

In order to fit the parameters $A_i$ and $B_i$ of the NP geometry
we first fit $\vert f^+\vert^2+\vert f^-\vert^2=2(f_{re}^2+f_{im}^2)$,
$Im(f^+f^{-*})=2f_{re}f_{im}$ and $t$ for each set of $(r,d)$ independently so
that the three $\lambda$ coincide with the three lowest eigenenergies.
In the case of $d> r +1$  the unperturbed energy of the
configuration B is lower than that of A and we found difficulty in fitting the
data. In these cases one should consider $A^\pm$ and B as the basis, but
since the  flux tubes of the basis B are not along the lattice link it is not
possible to specify the bases $A^\pm$. We omit these data from the analysis,
except $(r,d)=(1,2)$, where the difference in the unperturbed energy is
of the order of 0.05 and we can find parameters that fits the data
approximately.

We again fixed $E=0.296$ and $A_0=0.197$, and searched parameters to fit the
$\vert f^+\vert^2+\vert f^-\vert^2$ and $Im(f^+f^{-*})$ of the data sets
of $(r,d)$ that satisfy  $r\geq d-1$. We obtained $B_1=1.63(72)$,
$B_2=-0.12(1)$
and $A_4=6\pm 7$. $A_4$ consistent with zero means that the coherence of the
links on the twisted plaquette is weak. A fit with $A_4$ fixed to be zero
gives $B_1=5.3\pm 4.7$ and $B_2=-0.092(13)$. Although the error-bar of $B_1$
is large it is consistent with the previous 1.63(72). The coherence
on the bent plaquette parametrized by $B_2$ is relatively stable. It is
the main term that transforms among different symmetry states.  We remark
that $B_2$ in the NP case and $A_3$ in the TR case are both
related to the bent plaquette and the magnitudes are similar, while
$B_1$ and $A_1$ are both related to the rectangular plaquette but
the magnitude of $B_1$ is larger. In the latter case, not all the
4-quarks are sitting on the plaquette $\underline{B_2}$ and $\underline{B_3}$.
These plaquettes can be obtained by multiplying a plaquette to the base
$\underline{B_4}$, and by multiplying a plaquette to the base
$\underline{B_5}$,
respectively, on which all the 4-quarks are sitting.
Since a link in the model bases contains a staple-shaped link, due to the
fuzzing,  the effective area and the length of the perimeter of
the plaquette characterized by the parameter $B_2$ would be
modified. We cannot, however, estimate the correction and the parameter
$B_1$ is regarded as a phenomenological parameter.

The parameter $t$ that connects the bases $B^+$ and $B^-$ should tend
to 0 as r and d increase, and  we observed this tendency in the fitting.
 But for simplicity, we search an average value of the
parameter t that fits the three lowest eigenenergies of all the data of
$(r,d)$.
Taking $B_1=1.63, B_2=-0.12$ and $A_4=6.19$ and the error-bars of the
eigen-energies given by the lattice calculation data, we fitted t.
Due to relatively
large error-bars in the highest energy state, we sacrifice the accuracy of the
fit of the highest state. In fact in the highest state, more complicated
configurations could be involved and the fitting with three bases could be a
bad approximation. The value we obtained for $t$ is 0.045(10).
\end{enumerate}
\end{itemize}
%\end{section}

\section{Discussion and Conclusion}
We analyzed the four-quark energies in SU(2) calculated by lattice Monte
Carlo in a model which is inspired by the resonating group calculation
of hadron interactions. The transition matrix element between different
bases A and B is assumed to be proportional to the gluon overlap factor and
this factor is then parametrized by a product of an area
dependent and a perimeter dependent term. The perimeter dependent term
was introduced by considering transition between degenerate $Z_2$ symmetric
bases which in the large lattice or in the continuum limit would vanish
except the self energy term.
In the case of large square(LS), rectangular(R), tilted rectangular(TR),
linear(L) and quadrilateral(Q) configurations, we found reasonable parameters
for the perimeter dependent term which are characterized by the cubic
symmetry. In the non-planar(NP) case, we observed that it is necessary to take
into acount the gluonic degrees of freedom explicitly.  The base  B was split
into two hybrid configurations: $B^+$ and $B^-$. Using  A and  $B^\pm$ as the
bases, we parametrized the gluon overlap factor for the NP.

We treated, the error-bars
given by the lattice Monte Carlo as meaningful also for the small
plaquettes and checked whether a consistent parametrization for different
geometries is possible. We  simulated the deviation from an area
dependence by adding a perimeter dependent contribution.

The parameters for the perimeter dependent part are  $A_0, A_1, A_2, A_3, A_4,
B_1$ and $B_2$. $A_1'$ in the Q case is consistent with that of R case, and
$B_2$ in the NP case is consistent with $A_3$ in TR case. These coincidences
are not trivial
and are encouraging for any further analysis based on the cubic symmetry of
the flux tube geometry. $A_4$ is consistent with 0 and we could
well omit it. Therefore, there are 5 parameters for the perimeter
terms and one parameter for the area to fit the lattice Monte Carlo energy
spectra.

We expect that in the continuum limit $f$ is completely given by the area term
except for the self energy contribution. A comparison of the data at
$\beta=2.4$ and
$\beta=2.5$ for  the LS configuration suggests that the parameters of the
perimeter
term in fact decrease as the lattice constant $a$ becomes small.
Since the Monte Carlo
data contain lattice artefacts, in order to analyze the data
for a specific configuration we need to choose proper bases so that these
artefacts are eliminated. Since the artefacts are represented by the perimeter
term, its cubic symmetry property is essential for choosing the proper bases.

S.F thanks the Academy of Finland and the Japan Society for the Promotion of
Science for the support of the research from February 7 to April 8 of 1994.
He is also grateful to the Research Institute for Theoretical Physics of
the University of Helsinki for their kind hospitality.
%\end{section}

\newpage
\begin{figure}
\caption{The basic four-quark geometries of Large square(LS).
$\underline{A}$ is the configuration of the base A. $\underline{B_i}$ are the
configurations of the base B.  The linked $q\bar q$ of the bases A are
assigned 1-3 and 2-4, and those of the bases B are 1-4 and 2-3.}
\end{figure}
\begin{figure}
\caption{Same as Fig.1 but for Rectangle(R).}
\end{figure}
\begin{figure}
\caption{Same as Fig.1 but for Tilted rectangle(TR).}
\end{figure}
\begin{figure}
\caption{Same as Fig.1 but for Linear(L).}
\end{figure}
\begin{figure}
\caption{Same as Fig.1 but for Quadrilateral(Q).}
\end{figure}
\begin{figure}
\caption{Same as Fig.1 but for Non-planar(NP).}
\end{figure}

\newpage
\begin{table}[h]
\caption{$b_s, e$ and $v_0$
of the quark-antiquark potential }
\begin{center}
\begin{tabular}[h]{|r|r|r|r|}
\hline
%INSERT HERE THE DATA FILE.....START:
 & & &    \\ $\beta$ &
 $b_s$ &e   & $v_0$   \\ \hline
2.4 & 0.07169 & 0.245 & 0.550   \\
2.5 & 0.03781 & 0.215 & 0.523   \\
\hline
\end{tabular}
\end{center}
\end{table}

\end{document}